# Analytical formula and measurement technique for the built-in potential of practical diffused semiconductor junctions

Miron J. Cristea, *"Politehnica" University of Bucharest*, Romania
Tel :0040-21402-4915 Fax: 0040-214-300-555 E-mail: mcris@lydo.org

**Abstract**

Based on the Gauss law for the electric field, a new integral formula is deduced together with one of its possible applications, in the area of semiconductor junctions, specifically an analytical formula for the built-in potential of diffused semiconductor junctions. The measurement technique for the determination of the built-in potential is also described.

**1. Theory**

The Gauss law for the electric field over an electrically charged space region is [1]:

$$\frac{dE}{dx} = \frac{\rho(x)}{\varepsilon} \quad (1)$$

where $E$ is the electric field, $\rho$ is the electric space charge density and $\varepsilon$ the electrical permittivity. One-dimensional case is assumed, also linear, non-dispersive, isotropic material.

The electric field $E$ is connected with the electric potential $u$ in accordance with [1]:

$$E = -\frac{du}{dx} \quad (2)$$

By writing (1) as

$$dE = \frac{\rho(x)}{\varepsilon} dx \quad (3)$$

multiplying the equation by $x$, and taking into account that

$$d(xE) = xdE + Edx \quad (4)$$

the next equation is obtained:

$$\frac{\rho(x)}{\varepsilon}xdx = d(xE) - Edx \quad (5)$$

The integration of (5) over the space charge region (SCR) gives

$$\int_{SCR}\frac{x\rho(x)}{\varepsilon}dx = \int_{SCR}d(xE) - \int_{SCR}Edx \quad (6)$$

and taking (2) into account, then

$$\int_{SCR}\frac{x\rho(x)}{\varepsilon}dx = \int_{SCR}d(xE) + \int_{SCR}du \quad (7)$$

Since the electric field is zero at both ends of the SCR [2], the first term in the right hand of (7) vanishes, and the next equation is obtained:

$$\int_{SCR}\frac{x\rho(x)}{\varepsilon}dx = V_{SCR} \quad (8)$$

where $V_{SCR}$ is the total voltage drop across the space charge region.

**2. The formula of the built-in potential**

In the case of a semiconductor junction without external bias, $V_{SCR}$ equals the built-in potential of the junction $V_{bi}$:

$$V_{bi} = \int_{SCR}\frac{x\rho(x)}{\varepsilon}dx \quad (9)$$

For example, it is of interest to calculate the built-in potential of a diffused junction (i.e. having a Gaussian doping profile). For such junctions, the impurity profile, and in the depletion approximation, the space charge density (except for a factor of $q$ – the elementary electric charge) are given by:

$$N_d(x) = N_0 \exp\left(-\frac{x^2}{L_d^2}\right) \quad (10)$$

where $N_0$ is the surface concentration and $L_d$ is the technological diffusion length [1]

$$L_d = 2\sqrt{D_i \cdot t_d} \quad (11)$$

with $D_i$ - the doping impurity diffusion constant at certain diffusion temperature and $t_d$ the diffusion time. Inserting (10) this into eq.(9), the following relation is obtained:

$$V_{bi} = \frac{qN_0 L_d^2}{2\varepsilon}\left[1 - \exp\left(-\frac{W_{SC0}^2}{L_d^2}\right)\right] \qquad (12)$$

where Wsco is the depletion region width of the junction at no external bias.

### 3. Particular cases

If $W_{SC0} \ll L_d$, the formula simplifies to:

$$V_{bi} = \frac{qN_0 W_{SC0}^2}{2\varepsilon} \qquad (13)$$

Conversely, when $W_{SC0} \gg L_d$, (12) reduces to:

$$V_{bi} = \frac{qN_0 L_d^2}{2\varepsilon} \qquad (14)$$

Since $L_d$ is technologically determined, it is easy to discern between the limit cases (13) and (14).

### 4. Measuring the built-in potential

Since the zero-bias barrier capacitance per unit area of a junction is given by

$$C_{b0} = \frac{\varepsilon}{W_{SC0}} \qquad (15)$$

the barrier capacitance is measured in a standard circuit (that includes an AC-voltage generator), and then the built-in potential is calculated with the formula:

$$V_{bi} = \frac{qN_0 L_d^2}{2\varepsilon}\left[1 - \exp\left(-\frac{\varepsilon^2}{L_d^2 C_{b0}^2}\right)\right] \qquad (16)$$

If a calibrated DC-voltage source is also available, the measurement of the barrier capacitance can be realized at a determined reverse bias $V_R$ of the junction, and the built-in potential will be determined from the formula

$$V_R + V_{bi} = \frac{qN_0 L_d^2}{2\varepsilon}\left[1 - \exp\left(-\frac{\varepsilon^2}{L_d^2 C_b^2}\right)\right] \qquad (17)$$

where $C_b$ is the barrier capacitance corresponding to the reverse bias $V_R$.

**Conclusion**

Based on the Gauss law for the electric field, a new integral formula was deduced in this work, together with one of its possible applications, in the area of semiconductor junctions, specifically a new formula for the built-in potential of diffused semiconductor junctions. The measurement technique for the determination of the built-in potential was also given.